\documentclass[a4paper,10pt,notitlepage]{article}

\usepackage[latin1]{inputenc}
\usepackage{amssymb, bm}
\usepackage{graphicx}
\usepackage{lscape}
\usepackage{setspace}
\doublespace
\usepackage{lineno}
\linenumbers

\def\0{\phantom{0}}
\def\.{\phantom{.}}

\title{An optimized molecular model for ammonia}
\author{Bernhard Eckl, Jadran Vrabec\footnote{Tel.: +49-711/685-66107, Fax:
+49-711/685-66140, Email: vrabec@itt.uni-stuttgart.de}, and Hans Hasse}
\date{Institut f\"ur Technische Thermodynamik und Thermische Verfahrenstechnik, Universit\"at Stuttgart}

\begin{document}
\maketitle

% abstract
\begin{abstract}
An optimized molecular model for ammonia, which is based on a previous work of Krist\'of et al., Mol. Phys. 97 (1999) 1129--1137, is presented. Improvements are achieved by including data on geometry and electrostatics from \emph{ab initio} quantum mechanical calculations in a first model. Afterwards the parameters of the Lennard-Jones potential, modeling dispersive and repulsive interactions, are optimized to experimental vapor-liquid equilibrium data of pure ammonia. The resulting molecular model shows mean unsigned deviations to experiment of 0.7~\% in saturated liquid density, 1.6~\% in vapor pressure, and 2.7~\% in enthalpy of vaporization over the whole temperature range from triple point to critical point. This new molecular model is used to predict thermophysical properties in the liquid, vapor and supercritical region, which are in excellent agreement with a high precision equation of state, that was optimized to $1147$ experimental data sets. Furthermore, it is also capable to predict the radial distribution functions properly, while no structural information is used in the optimization procedure.
\end{abstract}

\vspace{5mm}

% main text
\section{Introduction}
\label{Einleitung}
Molecular modeling and simulation is a powerful tool for predicting thermophysical properties, that is becoming more accesible due to the ever increasing computing power and the progress of methods and simulation tools. For real life applications in process engineering reliable predictions are needed for a wide variety of properties \cite{Ungerer2006, Eckl2008, Eckl2007}.

The central role for that task is played by the molecular model, that determines all of them. Therefore, a balanced modeling procedure, i.e. selection of model type and parameterization, is crucial. Unfortunately, thermophysical properties usually depend on the model parameters in a highly non-linear fashion. So the development of new molecular models of technical quality is a time-consuming task. In this paper a procedure is proposed that uses information from \emph{ab initio} quantum mechanical calculations to accelerate the modeling process. As an example, ammonia is regarded here.

Ammonia is a well-known chemical intermediate, mostly used in fertilizer industries; another important application is its use as a refrigerant. Due to its simple symmetric structure and its strong intermolecular interactions it is also of high academic interest both experimentally and theoretically.

Different approaches can be found in the literature to construct an intermolecular potential for ammonia to be used in molecular simulation. Jorgensen and Ibrahim \cite{Jorgensen1980} as well as Hinchliffe et al. \cite{Hinchliffe1981} used experimental bond distances and angles to place their interaction sites. Jorgensen and Ibrahim fitted a 12-6-3 potential plus four partial charges to results from \emph{ab initio} quantum mechanical calculations, they derived for 250 orientations of the ammonia dimer using the STO-3G minimal basis set. To yield reasonable potential energies for liquid ammonia compared to experimental results, they had to scale their potential by a factor 1.26.

Hinchliffe et al. used a combination of exponential repulsion terms, an attractive Morse potential, and four partial charges to construct the intermolecular potential. The parameters were determined by fitting to a total of 61 points on the ammonia dimer energy surface at seven different orientations, which were calculated using the 6-31G* basis set. Hinchliffe et al. have pointed out, that the parameterization is ambiguous concerning the selection of dimer configurations and the used interaction potentials. Also the different models perform different well on various properties.

In a later work Impey and Klein \cite{Impey1984} reparameterized the molecular model by Hinchliffe et al. They switched to an "effective" pair potential using one Lennard-Jones (12-6) potential at the nitrogen nucleus site to describe the dispersive and repulsive interactions. The parameters were optimized to the radial distribution function $g_\mathrm{N-N}$ of liquid ammonia measured by Narten \cite{Narten1977}.

Krist\'of et al. \cite{Kristof1999} used this model to predict vapor-liquid equilibrium properties and found systematic deviations in both vapor pressure and saturated densities. So they decided to develop a completely new molecular model. Again they used experimental bond distances and angles to place the interaction sites. All further parameters of their model, i.e. the partial charges on all atoms and the parameters of the single Lennard-Jones (12-6) potential, were adjusted to vapor-liquid coexistence properties. With this model Krist\'of et al. reached a description of the vapor-liquid equilibrium (VLE) of ammonia of reasonable accuracy.

\begin{figure}[ht]
\includegraphics[scale=0.5]{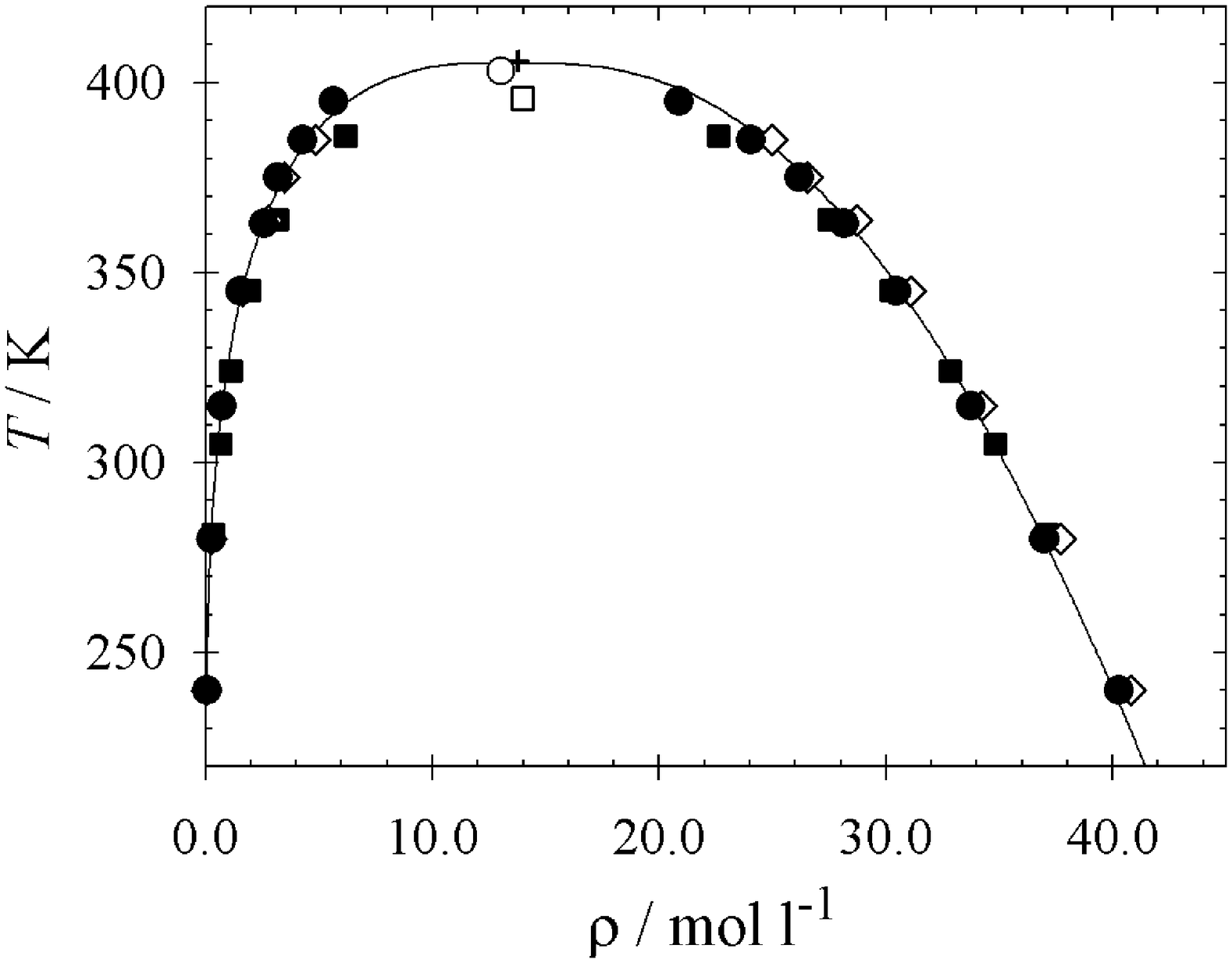}
\bigskip
\caption[Saturated densities of ammonia. Simulation results: {\Large $\bullet$}~new model, {\small $\blacksquare$}~model from Krist\'of et al., {\Large $\diamond$}~first model; {---}~reference EOS \cite{Tillner-Roth1993}, critical data from simulation: {\Large $\circ$}~new model, {\small $\square$}~model from Krist\'of et al.; $\bm{+}$~experimental critical point \cite{Mathews1972}.]{Eckl et al.\label{fig_vle_rho}} 
\end{figure}

\begin{figure}[ht]
\includegraphics[scale=0.5]{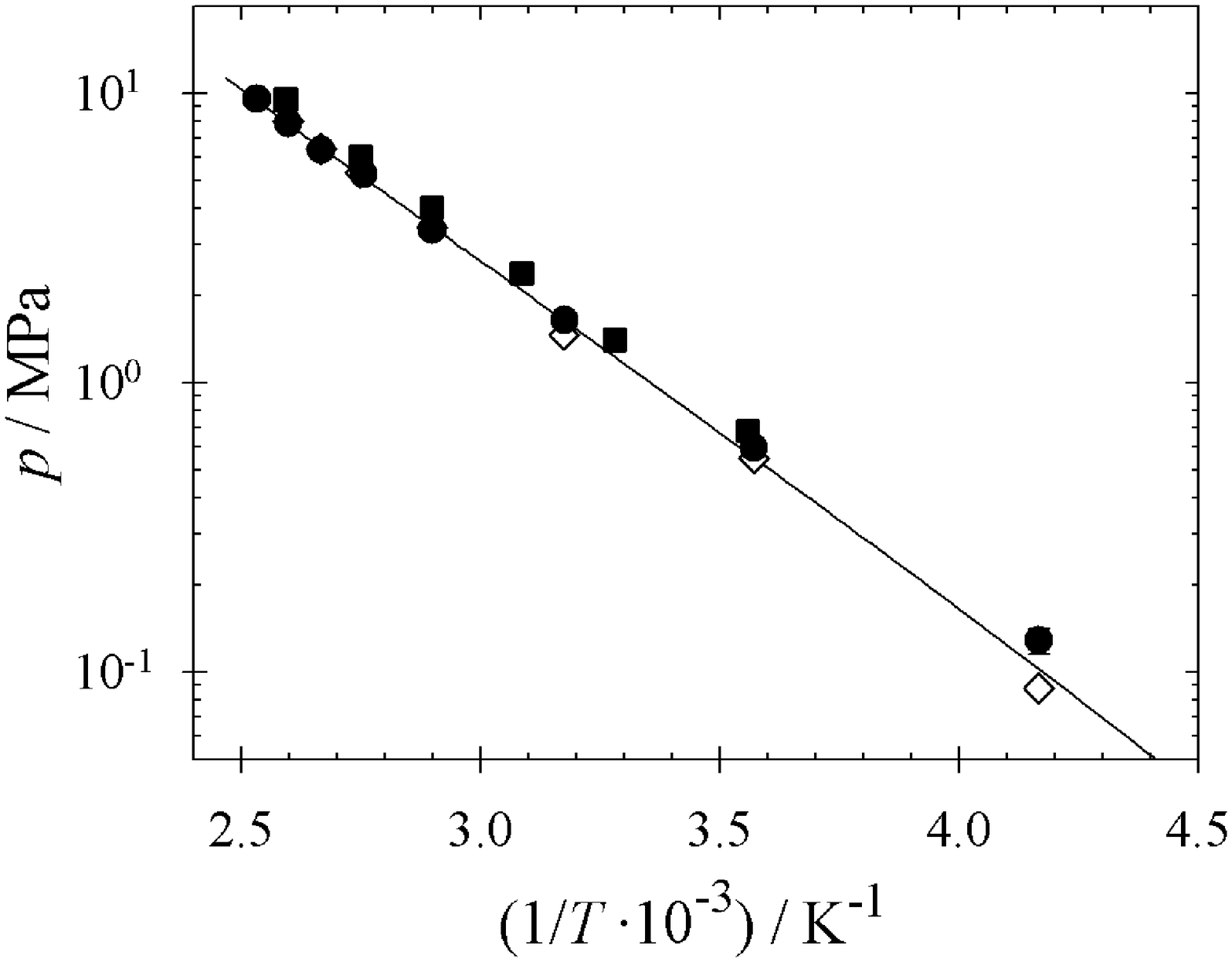}
\bigskip
\caption[Vapor pressure of ammonia. Simulation results: {\Large $\bullet$}~new model, {\small $\blacksquare$}~model from Krist\'of et al., {\Large $\diamond$}~first model; {---}~reference EOS \cite{Tillner-Roth1993}.]{Eckl et al.\label{fig_vle_p}}
\end{figure}

\begin{figure}[ht]
\includegraphics[scale=0.5]{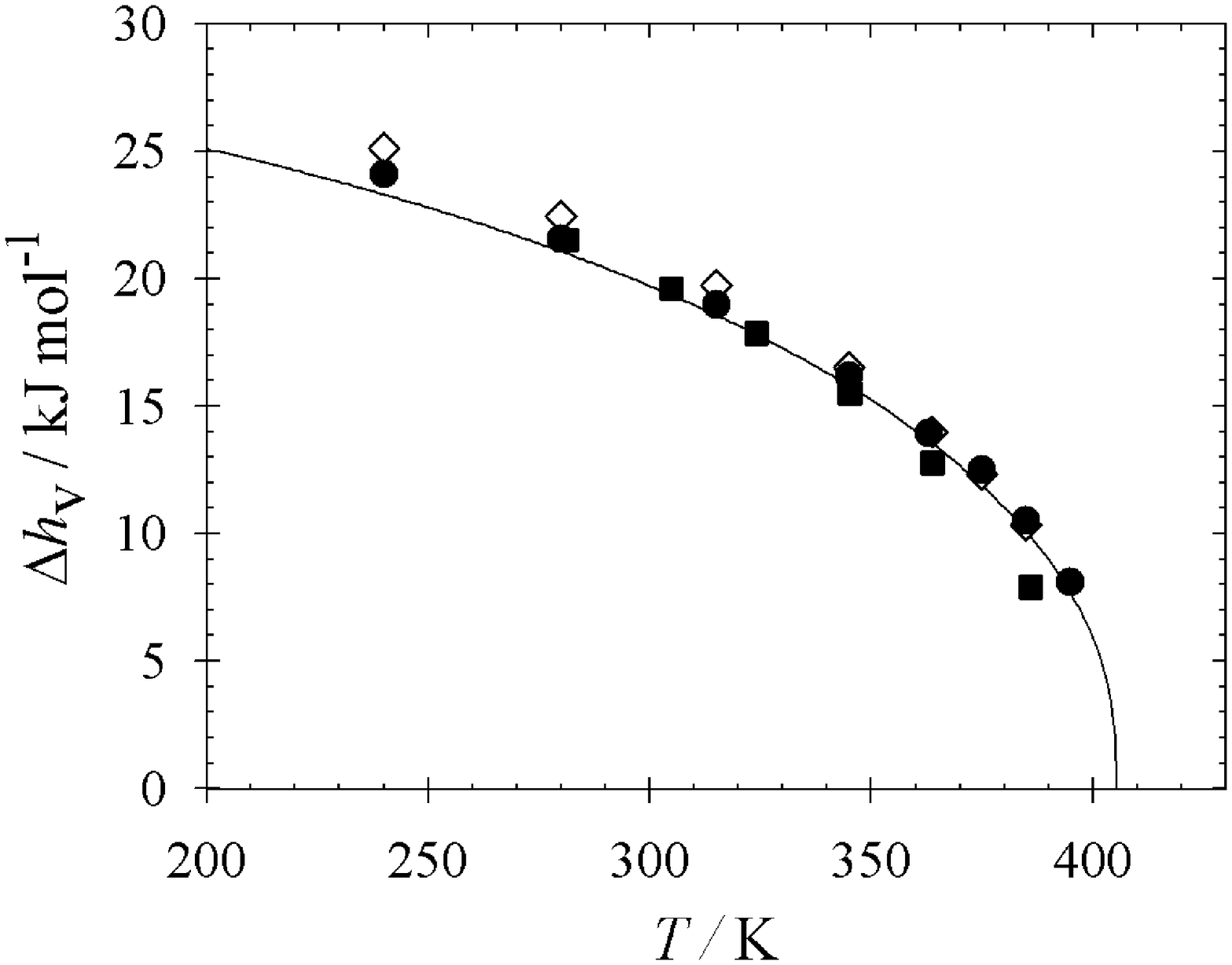}
\bigskip
\caption[Enthalpy of vaporization of ammonia. Simulation results: {\Large $\bullet$}~new model, {\small $\blacksquare$}~model from Krist\'of et al., {\Large $\diamond$}~first model; {---}~reference EOS \cite{Tillner-Roth1993}.]{Eckl et al.\label{fig_vle_dhv}}
\end{figure}

For their simulations, Krist\'of et al. used the Gibbs ensemble Monte Carlo (GEMC) technique \cite{Pana1987, Pana1988} with an extension to the $NpH$ ensemble \cite{Kristof1996, Kristof1997}. This methods have some difficulties simulating strongly interacting fluids, yielding to relatively large statistical uncertainties. When applying our methods for the simulation of VLE, leading to much smaller statistical errors, we get results slightly outside the error bars of Krist\'of et al. Also systematic deviations to the experimental vapor-liquid properties are seen, especially in the vapor pressure and critical temperature, cf. Figures~\ref{fig_vle_rho} to \ref{fig_vle_dhv}.

In the present work a new molecular model for ammonia is proposed. This model is based on the work of Krist\'of et al. and improved by including data on geometry and electrostatics from \emph{ab initio} quantum mechanical calculations.

The paper is structured as follows: Initially, a procedure is proposed for the development of a preliminary molecular model. This model, called \emph{first model} in the following, is then adjusted to experimental VLE data until a desired quality is reached. The resulting model, denoted as \emph{new model} in the following, is used afterwards to predict thermal and caloric properties apart from the phase coexistence as well as structural properties.

\section{Selection of Model Type and Parameterization}
The modeling philosophy followed here is to keep the molecular model as simple as possible. Therefore, the molecule is assumed rigid and non-polarizable, i.e. a single state-independent set of parameters is used. Hydrogen atoms are not modeled explicitely, a united-atom approach is used.

For both present models, a single Lennard-Jones potential was assumed to describe the dispersive and repulsive interactions. The electrostatic interactions as well as hydrogen bonding were modeled by a total of four partial charges. This modeling approach was found to be appropriate for other hydrogen bonding fluids like methanol \cite{Schnabel2007}, ethanol \cite{Schnabel2005}, and formic acid \cite{Schnabel2007a} and was also followed by Impey and Klein \cite{Impey1984} and Krist\'of et al. \cite{Kristof1999} for ammonia.

Thus, the potential energy $u_{ij}$ between two ammonia molecules $i$ and $j$ is given by

\noindent
\begin{equation}\label{eq_potential}
  u_{ij} (r_{ij}) = 4 \varepsilon \left[ \left( \frac{\sigma}{r_{ij}} \right) ^{12} - \left( \frac{\sigma}{r_{ij}} \right) ^{6} \right] + \sum_{a=1}^4\sum_{b=1}^4 \frac{q_{ia}q_{jb}}{4\pi\epsilon_0r_{ijab}},
\end{equation}

\noindent where $a$ is the site index of charges on molecule $i$ and $b$ the site index of charges on molecule $j$, respectively. The site-site distances between molecules $i$ and $j$ are denoted by $r_{ij}$ for the single Lennard-Jones potential and $r_{ijab}$ for the four partial charges, respectively. $\sigma$ and $\varepsilon$ are the Lennard-Jones size and energy parameters, while $q_{ia}$ and $q_{jb}$ are the partial charges located at the sites $a$ and $b$ on the molecules $i$ and $j$, respectively. Finally, $\epsilon_0$ denotes the permittivity of the vacuum.

To keep the modeling procedure as independent as possible from the availability of specific information, no experimental bond lengths or angles were used here in contrast to \cite{Jorgensen1980, Hinchliffe1981, Impey1984, Kristof1999}. Instead, the nucleus positions were calculated by the means of quantum mechanics, where the software package GAMESS (US) \cite{Schmidt1993} was used. A geometry optimization was performed on the Hartree-Fock, i.e.~self-consistent field (SCF), level using the basis set 6-31G, which is a split-valence orbital basis set without polarizable terms. The nucleus positions from this \emph{ab initio} calculation were directly used to specifiy the positions of the five interaction sites. At the nitrogen nucleus site and at each of the hydrogen nucleus sites a partial charge was placed. The Lennard-Jones site conincides with the nitrogen nucleus position, cf. Table~\ref{tab_Parameter}.

To obtain the magnitude of the partial charges, another subsequent quantum mechanical calculation was performed. It was done on M\o{}ller-Plesset 2 level using the polarizable basis set 6-311G(d,p) and the geometry from the previous step. By default, quantum mechanical calculations are performed on a single molecule of interest in vacuum. It is widely known, that the gas phase dipolar moments significantly differ from the dipole moment in the liquid state. As it was seen from former work \cite{Vrabec2001, Stoll2003}, molecular models yield better results on VLE properties, when a "liquid-like" dipolar moment is applied. Therefore, the single molecule was calculated within a dielectric cavity utilizing the COSMO (COnducter like Screening MOdel) method \cite{Baldridge1997} to mimic the liquid state. The partial charges were chosen to yield the resulting dipole moment of $1.94$~Debye, the parameters are given in Table~\ref{tab_Parameter}.

The first model combines this electrostatics with the Lennard-Jones parameters of Krist\'of et al \cite{Kristof1999}, so no additional experimental data was used. To achieve an optimized new model, the two Lennard-Jones parameters $\sigma$ and $\varepsilon$ were adjusted to experimental saturated liquid density, vapor pressure, and enthalpy of vaporization using a Newton scheme as proposed by Stoll \cite{Stoll2005}. These properties were chosen for the adjustment as they all represent major characteristics of the fluid region. Furthermore, they are relatively easy to be measured and are available for many components of technical interest.

The applied optimization method has many similarities with the one proposed by Ungerer et al. \cite{Ungerer1999} and later on modified by Bourasseau et al. \cite{Bourasseau2003}. It relies on a least-square minimiztion of a weighted fitness function $\mathcal{F}$ that quantifies the devitions of simulation results from a given molecular model compared to experimental data. The weighted fitness function writes as

\noindent
\begin{equation}\label{eq_fitness}
  \mathcal{F} = \frac{1}{d} \sum_{i=1}^d \frac{1}{(\delta A_{i,\mathrm{sim}})^2} (A_{i,\mathrm{sim}}(\bm{M}_0) - A_{i,\mathrm{exp}})^2~,
\end{equation}

\noindent wherein the $n$-dimensional vector $\bm{M}_0 = (m_{0,1}, ..., m_{0,n})$ is a short-cut notation for the set of $n$ model parameters $m_{0,1}, ..., m_{0,n}$ to be optimized. The deviations of results from simulation $A_{i,\mathrm{sim}}$ to experimental data $A_{i,\mathrm{exp}}$ are weighted with the expected simulation uncertainties $\delta A_{i,\mathrm{sim}}$. Equation~(\ref{eq_fitness}) allows simultaneous adjustment of the model parameters to different thermophysical properties $A$ (saturated liquid densities $\rho'$, vapor pressures $p^\sigma$, and enthalpies of vaporization $\Delta h_\mathrm{v}$ at various temperatures in the present work).

The unknown functional dependence of the property $A$ on the model parameters is approximated by a first order Taylor series developed in the vicinity of the initial parameter set $\bm{M}_0$

\noindent
\begin{equation}
  A_{i,\mathrm{sim}}(\bm{M}_\mathrm{new}) = A_{i,\mathrm{sim}}(\bm{M}_0) + \sum_{j=1}^n \frac{\partial A_{i,\mathrm{sim}}}{\partial m_j} \cdot (m_{\mathrm{new},j} - m_{0,j})~.
\end{equation}

\noindent Therein, the partial derivatives of $A_i$ with respect to each model parameter $m_j$, i.e. the sensitivities, are calculated from difference quotients

\noindent
\begin{equation}
  \frac{\partial A_{i,\mathrm{sim}}}{\partial m_j} \approx \frac{A_{i,\mathrm{sim}}(m_{0,1}, ..., m_{0,j}+\Delta m_j, ..., m_{0,n}) - A_{i,\mathrm{sim}}(m_{0,1}, ..., m_{0,j}, ..., m_{0,n})}{\Delta m_j}~.
\end{equation}

Assuming a sound choice of the model parameter variations $\Delta m_j$, i.e. small enough to ensure linearity and large enough to yield differences in the simulation results significantly above the statistical uncertainties, this method allows a step-wise optimization of the molecular model by minimization of the fitness function $\mathcal{F}$. Experience shows that an optimized set of model parameters can be found within a few steps when starting from a reasonable initial model.

\section{Results and Discussion}
\subsection{Vapor-Liquid Equilibria}
VLE results for the new model are compared to data obtained from a reference quality equation of state (EOS) \cite{Tillner-Roth1993} in Figures~\ref{fig_vle_rho} to \ref{fig_vle_dhv}. These figures also include the results, that we calculated using the first model and the model from Krist\'of et al. \cite{Kristof1999}. The present numerical simulation results together with experimental data \cite{Tillner-Roth1993} are given in Table~\ref{tab_vle}, technical simulation details are given in the appendix.

The reference EOS \cite{Tillner-Roth1993} used for adjustment and comparison here, was optimized to $1147$ experimental data sets. It is based on two older EOS from the late nineteen seventies \cite{Haar1978, Ahrendts1979} and also recommended by the NIST within their reference EOS database REFPROP \cite{REFPROP}. The proposed uncertainties of the equation of state are 0.2~\% in density, 2~\% in heat capacity, and 2~\% in the speed of sound, except in the critical region. The uncertainty in vapor pressure is 0.2~\%.

The model of Krist\'of et al. shows noticeable deviations from experimental data. The mean unsigned errors over the range of VLE are 1.9~\% in saturated liquid density, 13~\% in vapor pressure and 5.1~\% in enthalpy of vaporization. Even without any further adjustment to experimental data a better description was found using the first model. The deviations between simulation results and reference EOS are 1.5~\% in saturated liquid density, 10.4~\% in vapor pressure and 5.1~\% in enthalpy of vaporization.

\begin{figure}[ht]
\includegraphics[scale=0.5]{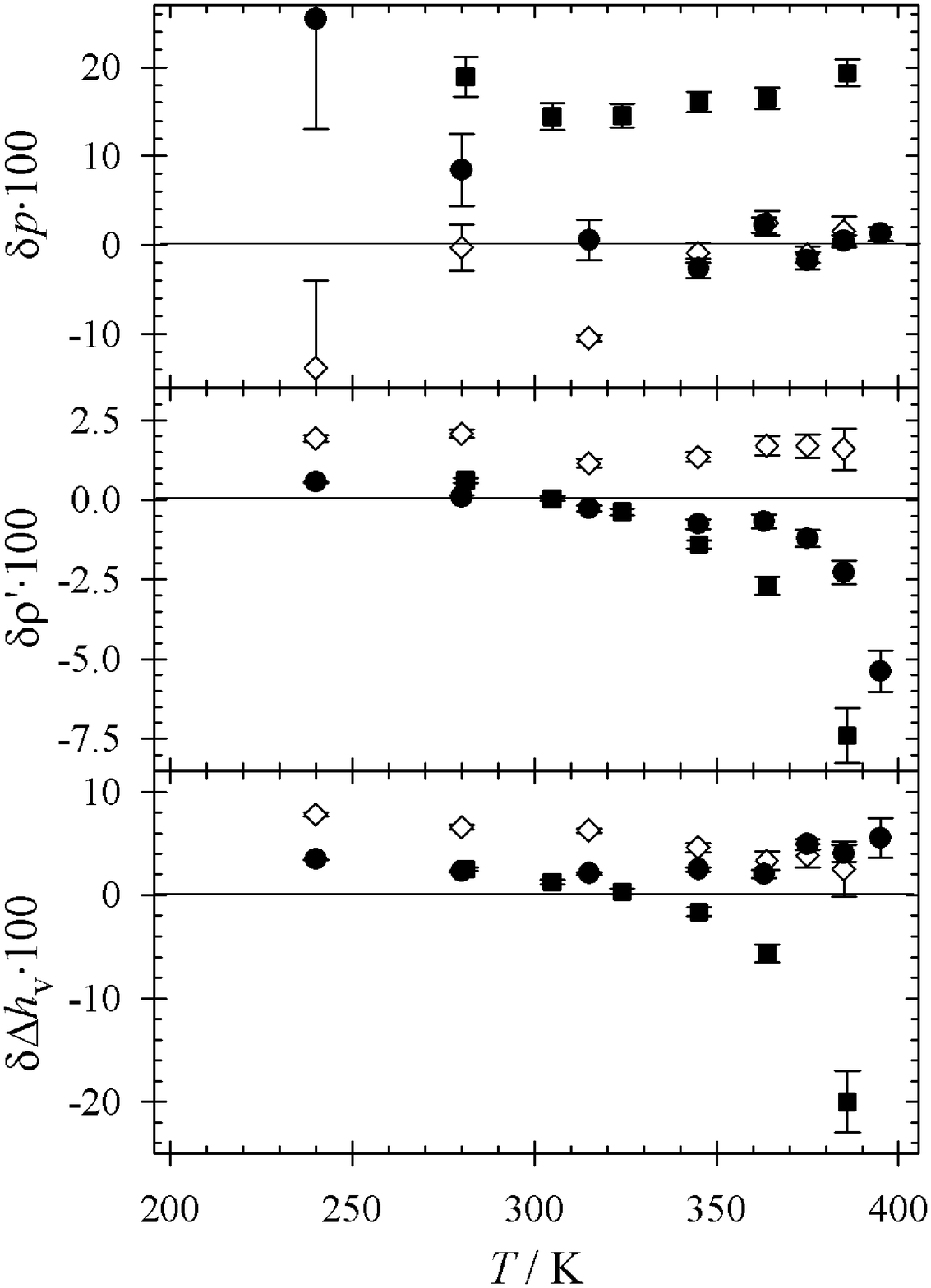}
\bigskip
\caption[Relative deviations of vapor-liquid equilibrium properties between simulation and reference EOS \cite{Tillner-Roth1993} ($\delta z = (z_{\mathrm{sim}} - z_{\mathrm{eos}}) / z_{\mathrm{eos}}$): {\Large $\bullet$}~new model, {\small $\blacksquare$}~model from Krist\'of et al., {\Large $\diamond$}~first model. Top: vapor pressure, center: saturated liquid density, bottom: enthalpy of vaporization.]{Eckl et al.\label{fig_vle_dev}}
\end{figure}

With the new model a significant improvement is achieved compared to the model from Krist\'of et al. The description of the experimental VLE is very good, the mean unsigned deviations in saturated liquid density, vapor pressure and enthalpy of vaporization are 0.7, 1.6, and 2.7~\%, respectively. Only at low temperatures, in the range of ambient pressure, a slightly worse description
of the vapor pressure compared to the first model is observed. In Figure~\ref{fig_vle_dev} the relative deviations of the new model, the model from Krist\'of et al., and the first model are shown in the whole range of the VLE starting from triple point to critical point.

Mathews \cite{Mathews1972} gives experimental critical values of temperature, density and pressure for ammonia: $T_\mathrm{c}=$405.65~K, $\rho_\mathrm{c}=$13.8~mol/l, and $p_\mathrm{c}=$11.28~MPa. Following the procedure suggested by Lotfi et al. \cite{Lotfi1992} the critical properties $T_\mathrm{c}=$395.82~K, $\rho_\mathrm{c}=$14.0~mol/l, and $p_\mathrm{c}=$11.26~MPa for the model of Krist\'of et al. were calculated, where the critical temperature is underestimated by 2.4~\%. For the first model $T_\mathrm{c}=$403.99~K, $\rho_\mathrm{c}=$14.1~mol/l, and $p_\mathrm{c}=$11.67~MPa were obtained and for the new model $T_\mathrm{c}=$402.21~K, $\rho_\mathrm{c}=$13.4~mol/l, and $p_\mathrm{c}=$10.52~MPa. The latter two give reasonable results for the critical temperature, while the new model underpredicts the critical pressure slightly.

\subsection{Homogeneous Region}
In many technical applications thermodynamic properties in the homogeneous fluid region apart from the VLE are needed. Thus, the new molecular model was tested on its predictive capabilities in these states.

\begin{figure}[ht]
\includegraphics[scale=0.5]{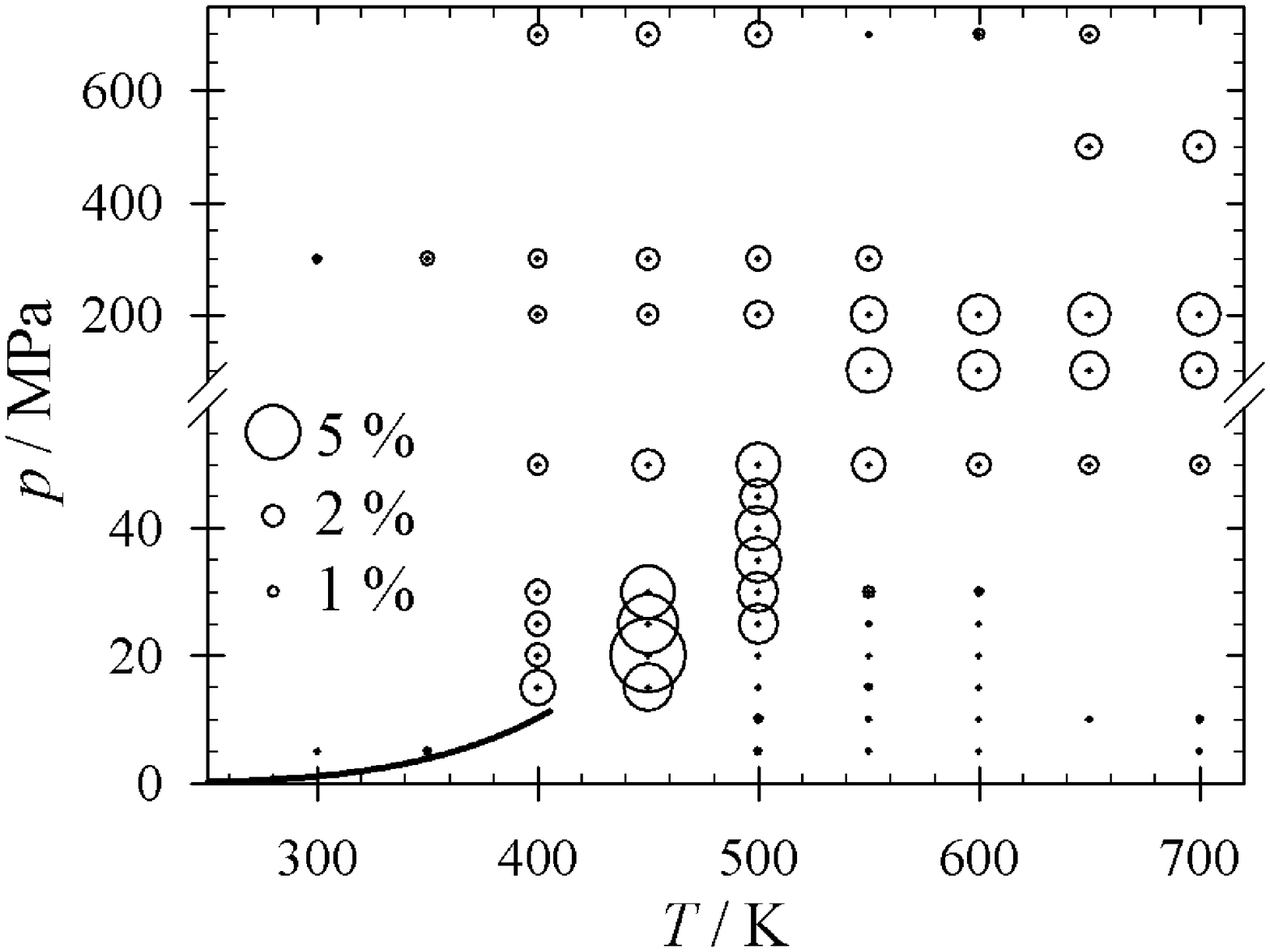}
\bigskip
\caption[Relative deviations for the density between simulation and reference EOS \cite{Tillner-Roth1993} ($\delta \rho = (\rho_{\mathrm{sim}} - \rho_{\mathrm{eos}}) / \rho_{\mathrm{eos}}$) in the homogeneous region: {\Large $\circ$}~simulation data of new model, {---}~vapor pressure curve. The size of the bubbles denotes the relative deviation as indicated in the plot.]{Eckl et al.\label{fig_hom_rho}}
\end{figure}

Thermal and caloric properties were predicted with the new model in the homogenous liquid, vapor and supercritical fluid region. In total, 70 state points were regarded, covering a large range of states with temperatures up to 700~K and pressures up to 700~MPa. In Figure~\ref{fig_hom_rho}, relative deviations between simulation and reference EOS \cite{Tillner-Roth1993} in terms of density are shown. The deviations are typically below 3~\% with the exception of the extended critical region, where a maximum deviation of 6.8~\% is found.

\begin{figure}[ht]
\includegraphics[scale=0.5]{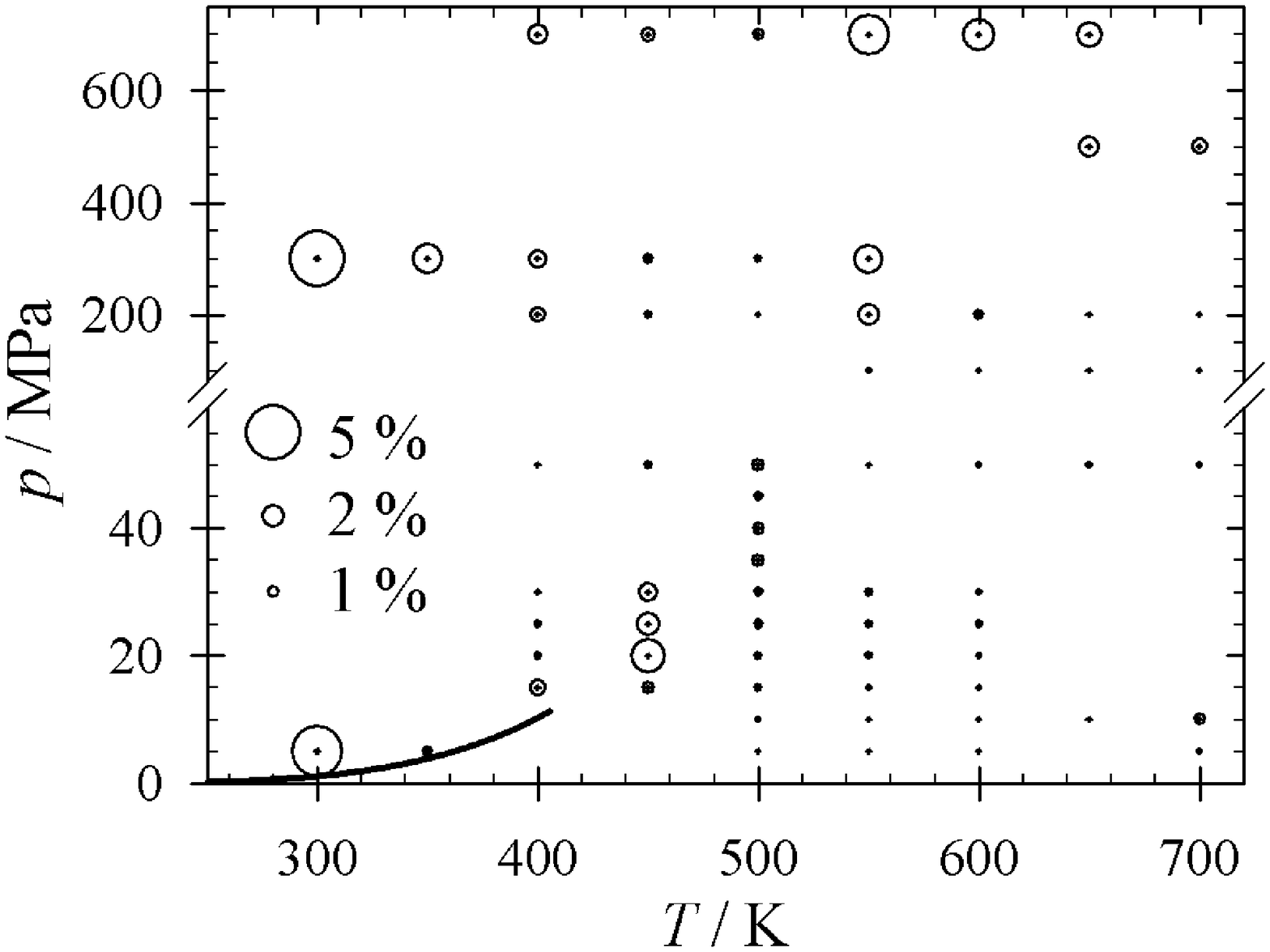}
\bigskip
\caption[Relative deviations for the enthalpy between simulation and reference EOS \cite{Tillner-Roth1993} ($\delta h = (h_{\mathrm{sim}} - h_{\mathrm{eos}}) / h_{\mathrm{eos}}$) in the homogeneous region: {\Large $\circ$}~simulation data of new model, {---}~vapor pressure curve. The size of the bubbles denotes the relative deviation as indicated in the plot.]{Eckl et al.\label{fig_hom_h}}
\end{figure}

Figure~\ref{fig_hom_h} presents relative deviations in terms of enthalpy between simulation and reference EOS \cite{Tillner-Roth1993}. In this case deviations are very low for low pressures and high temperatures (below 1--2~\%). Typical deviations in the other cases are below 5~\%.

These results confirm the modeling procedure. By adjustment to VLE data only, quantitatively correct predictions in most of the technically important fluid region can be obtained.

\subsection{Second Virial Coefficient}
The virial expansion gives an equation of state for low density gases. For ammonia it is a good approximation for gaseous states below 0.1~MPa with a maximum error of 2.5~\%. Starting from the nineteen thirties it was shown, that the virial coefficients can nowadays easily be derived from the intermolecular potential \cite{Mayer1937, Mayer1939, Mayer1940}. The second virial coefficient is related to the molecular model by~\cite{Gray1984}

\noindent
\begin{equation}
  B = -2\pi \int_0^{\infty} \left\langle \exp\left( -\frac{u_{ij}(r_{ij},\bm{\omega}_i,\bm{\omega}_j)}{k_{\mathrm B}T}\right) -1\right\rangle_{\bm{\omega}_i,\bm{\omega}_j} r_{ij}^2 \mathrm{d}r_{ij},
\end{equation}

\noindent where $u_{ij}(r_{ij},\bm{\omega}_i,\bm{\omega}_j)$ is the interaction energy between two molecules $i$ and $j$, cf. Equation~(\ref{eq_potential}). $k_{\mathrm B}$ denotes Boltzmann's constant and the $\langle\rangle$ brackets indicate an average over the orientations $\bm{\omega}_i$ and $\bm{\omega}_j$ of the two molecules separated by the center of mass distance $r_{ij}$.

The second virial coefficient was calculated here by evaluating Mayer's $f$-function at $363$ radii from $2.4$ to $8$~\r{A}, averaging over $500^2$ random orientations at each radius. The random orientations were generated using a modified Monte Carlo scheme \cite{Mountain2005, Eckl2008}. A cut-off correction was applied for distances larger than $8$~\r{A} for the LJ potential \cite{Allen1987}. The electrostatic interactions need no long-range correction as they vanish by angle averaging.

\begin{figure}[ht]
\includegraphics[scale=0.5]{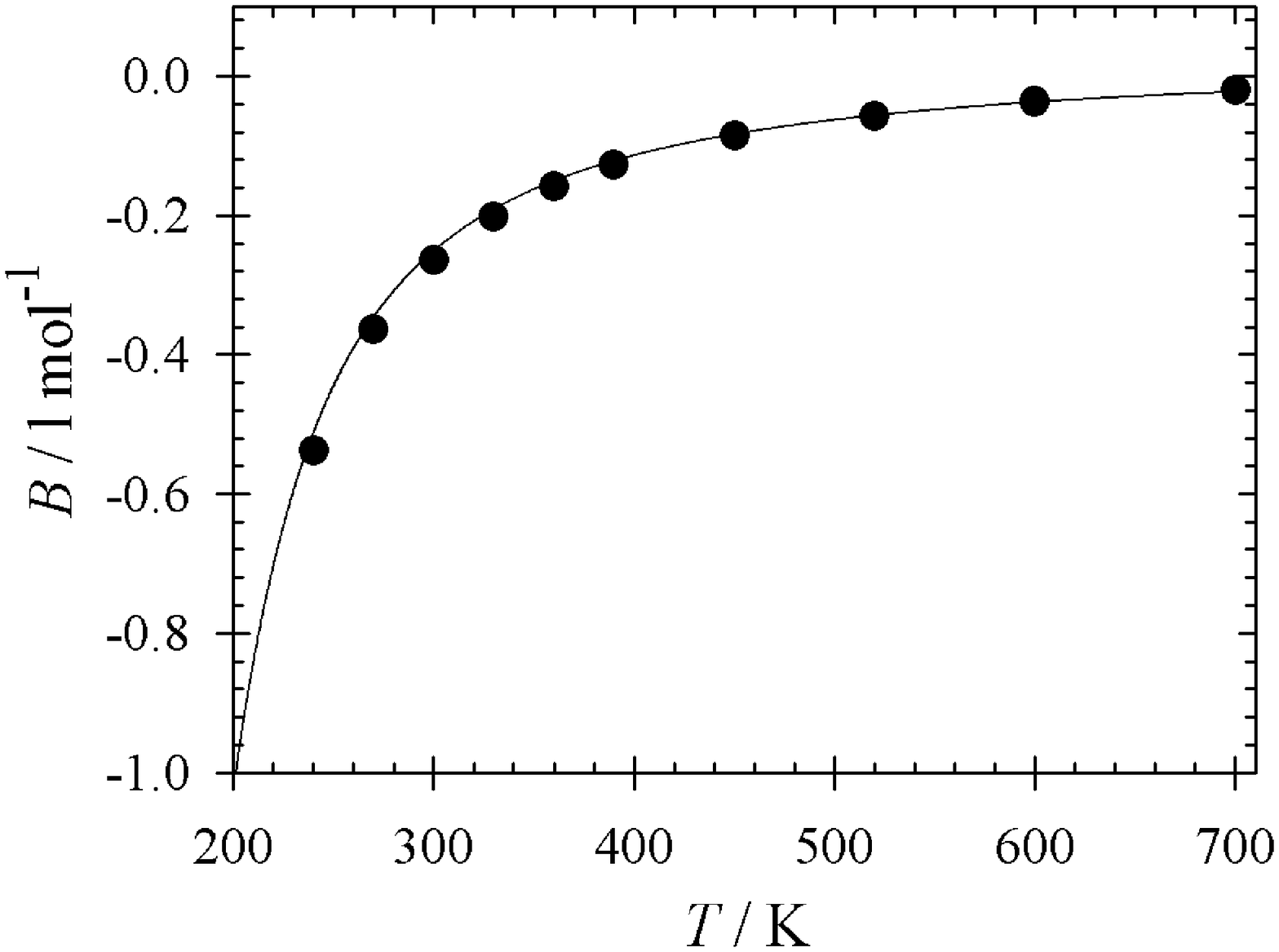}
\bigskip
\caption[Second virial coefficient: {\Large $\bullet$}~simulation data of new model, {---}~reference EOS \cite{Tillner-Roth1993}.]{Eckl et al.\label{fig_2vk}}
\end{figure}

Figure~\ref{fig_2vk} shows the second virial coefficient predicted by the new model is shown in comparison to the reference EOS. An excellent agreement was found over the full temperature range with a maximum deviation of $-4.3$~\% at $300$~K.

\subsection{Structural Quantities}
\begin{figure}[ht]
\includegraphics[scale=0.5]{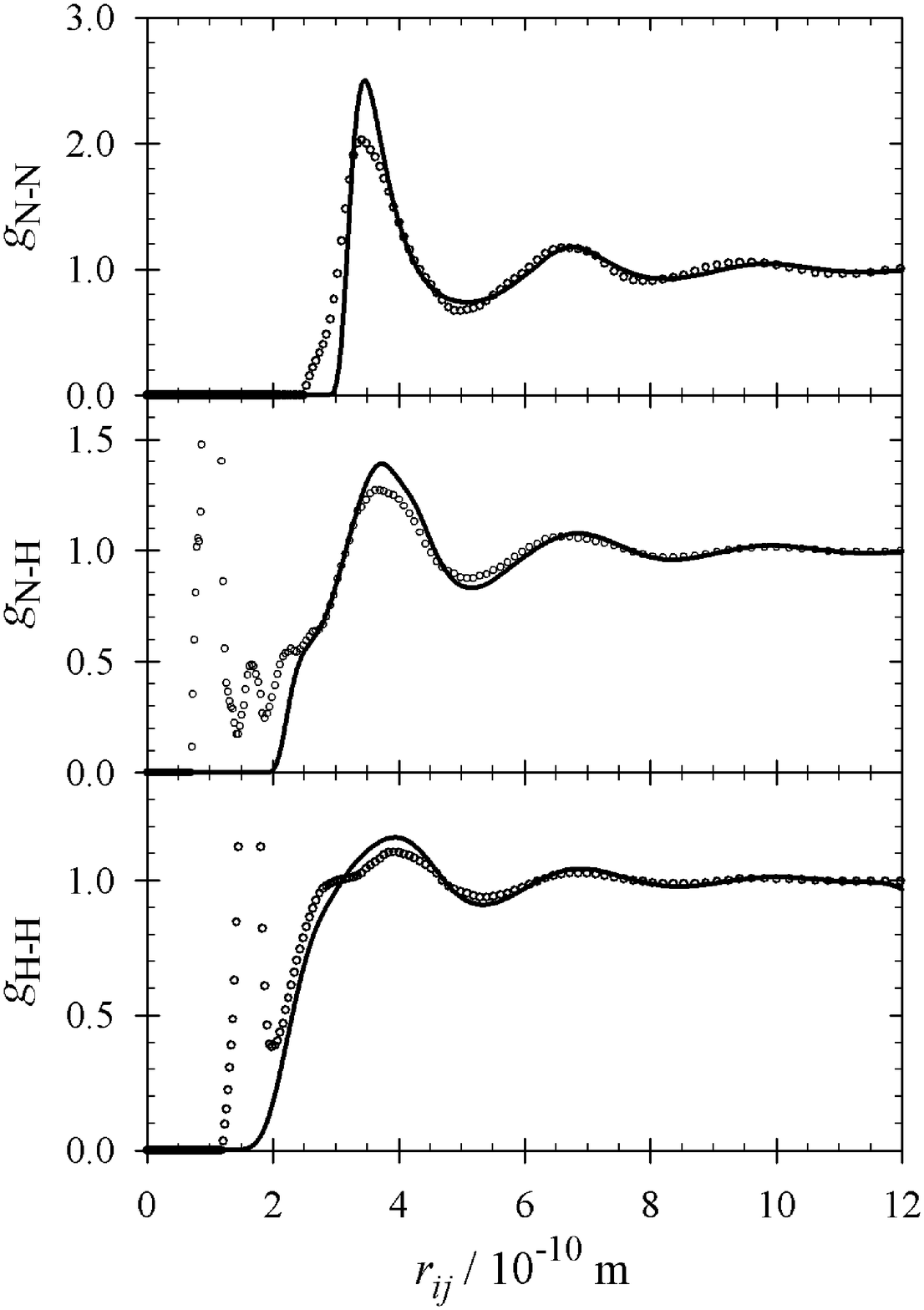}
\bigskip
\caption[Partition functions of ammonia: {---}~simulation data of new model, {\small $\circ$}~experimental data \cite{Ricci1995}.]{Eckl et al.\label{fig_rdf}}
\end{figure}

Due to its scientific and technical importance, experimental data on the microscopic structure of liquid ammonia are available. Narten \cite{Narten1977} and Ricci et al. \cite{Ricci1995} applied X-ray and neutron diffraction, respectively. The results from Ricci et al. show a smoother gradient and are available for all three types of atom-atom pair correlations, namely nitrogen-nitrogen (N-N), nitrogen-hydrogen (N-H), and hydro\-gen-\-hydro\-gen (H-H), thus they were used for comparison here. In Figure~\ref{fig_rdf}, these experimental radial distribution functions for liquid ammonia at 273.15~K and 0.483~MPa are compared to present predictive simulation data based on the new model.

It is found that these structural properties are in very good agreement, although no adjustment was done with regard to structural properties. The atom-atom distance of the first three layers is predicted correctly, while only minor overshootings in the first peak are found. Please note, that the first peak of experiment in $g_\mathrm{N-H}$ and $g_\mathrm{H-H}$ show intramolecular pair correlations, which are not calculated in the simulation.

In the experimental radial distribution function $g_\mathrm{N-H}$ the hydrogen bonding of ammonia can be seen at 2--2.5~\r{A}. Due to the simplified approximation by off-centric partial charges, the molecular model is not capable to describe this effect completely. But even with this simple model a small shoulder at 2.5~\r{A} is obtained.

\section{Conclusion}
A new molecular model is proposed for ammonia. This model was developed using a new modeling procedure, which speeds up the modeling process and can be applied on arbitrary molecules. The interaction sites are located according to atom positions resulting from \emph{ab initio} quantum mechanical calculations. Also the electrostatic interactions, here in form of partial charges, are parameterized according to high-level \emph{ab initio} quantum mechanical results. The latter are obtained by calculations within a dielectric continuum to mimic the (stronger) interactions in the liquid phase. The partial charges for the present ammonia model are specified to yield the same dipole moment as quantum mechanics. The Lennard-Jones parameters were adjusted to VLE data, namely vapor pressure, saturated liquid density, and enthalpy of vaporization.

A description of the VLE of ammonia was reached within relative deviations of a few percents. Next to this, covering a large region of states, a good prediction of both thermal and caloric properties apart from the VLE was found compared to a reference EOS \cite{Tillner-Roth1993}.

Predicted structural quantities, i.e. radial distribution functions in the liquid state, are in very good agreement to experimental neutron diffraction data. This shows, that molecular models adjusted to macroscopic thermodynamic properties also give reasonable results on microscopic properties. Note that this is not true vice versa in most cases. With the present model a similar quality in describing the atomic radial distribution functions as Impey and Klein \cite{Impey1984} is gained, while the macroscopic properties like vapor pressure differ considerably. So the latter can be seen as the more demanding criteria.

\section{Acknowledgment}
The authors gratefully acknowledge financial support by Deutsche Forschungsgemeinschaft, Schwerpunktprogramm 1155 "Molecular Modeling and Simulation in Process Engineering". The simulations were performed on the national super computer NEC SX-8 at the High Performance Computing Center Stuttgart (HLRS) under the grant MMHBF.
We also want to thank Mr. Xijun Fu for setting up and running the simulations shown in Figures~\ref{fig_hom_rho} and \ref{fig_hom_h}.

% \appendix
\section{Appendix}
\label{Appendix}
The Grand Equilibrium method \cite{Vrabec2002} was used to calculate VLE data at eight temperatures from 240 to 395~K during the optimization process. At each temperature for the liquid, molecular dynamics simulations were performed in the $NpT$ ensemble using isokinetic velocity scaling \cite{Allen1987} and Anderson's barostat \cite{Anderson1980}. There, the number of molecules is $864$ and the time step was $0.58$~fs except for the lowest temperature, where $1372$ molecules and a time step of $0.44$~fs were used. The initial configuration was a face centered cubic lattice, the fluid was equilibrated over $120~000$ time steps with the first $20~000$ time steps in the canonical ($NVT$) ensemble. The production run went over $300~000$ time steps ($400~000$ for 240~K) with a membrane mass of $10^9$ kg/m$^4$. Widom's insertion method \cite{Widom1963} was used to calculate the chemical potential by inserting up to $4~000$ test molecules every production time step.

At the lowest two temperatures additional Monte Carlo simulations were performed in the $NpT$ ensemble for the liquid. There, the chemical potential of liquid ammonia was calculated by the gradual insertion method \cite{Vrabec2002a}. The number of molecules was $500$. Starting from a face centered cubic lattice, $15~000$ Monte Carlo cycles were performed for equilibration and $50~000$ for production, each cycle containing $500$ displacement moves, $500$ rotation moves, and $1$ volume move. Every $50$ cycles $5000$ fluctuating state change moves, $5000$ fluctuating particle translation/rotation moves, and $25000$ biased particle translation/rotation moves were performed, to measure the chemical potential. These computationally demanding simulations yield the chemical potential in the dense and strong interacting liquid with high accuracy, leading to small uncertainties in the VLE.

For the corresponding vapor, Monte Carlo simulations in the pseudo-$\mu VT$ ensemble were performed. The simulation volume was adjusted to lead to an average number of $500$ molecules in the vapor phase. After $1~000$ initial $NVT$ Monte Carlo cycles, starting from a face centered cubic lattice, $10~000$ equilibration cycles in the pseudo-$\mu VT$ ensemble were performed. The length of the production run was $50~000$ cycles. One cycle is defined here to be a number of attempts to displace and rotate molecules equal to the actual number of molecules plus three insertion and three deletion attempts.

The cut-off radius was set to $17.5$~\r{A} throughout and a center of mass cut-off scheme was employed. Lennard-Jones long-range interactions beyond the cut-off radius were corrected as proposed in \cite{Allen1987}. Electrostatic interactions were approximated by a resulting molecular dipole and corrected using the reaction field method \cite{Allen1987}. Statistical uncertainties in the simulated values were estimated by a block averaging method \cite{Flyvbjerg1989}.

For the simulations in the homogeneous region, molecular dynamics simulations were performed with the same technical parameters as used for the saturated liquid runs.

For the radial distribution functions a molecular dynamics simulation was performed with $500$ molecules. Intermolecular site-site distances were divided in $200$ slabs from 0 to 13.5~\r{A} and summed up for $50~000$ time steps.

\clearpage

% \section{}
% \label{}

% tables
\newpage
\begin{table}[h]
\noindent
\caption{Parameters of the new ammonia model. The electronic charge is $e = 1.6021 \cdot 10^{-19}$~C.}
\label{tab_Parameter}

\medskip
\begin{center}
\begin{tabular}{lcccccc} \hline\hline
Interaction & $x$         & $y$         & $z$         & $\sigma$ & $\varepsilon/k_\mathrm{B}$ & $q$ \\
Site        & \r{A}       & \r{A}       & \r{A}       & \r{A}    & K             & $e$      \\ \hline
N           & \.0.0\0\0\0 & \.0.0\0\0\0 & \.0.0757    & 3.376    & 182.9         & -0.9993  \\
H(1)        & \.0.9347    & \.0.0\0\0\0 & -0.3164     & ---      & ---           & \.0.3331 \\
H(2)        & -0.4673     & \.0.8095    & -0.3164     & ---      & ---           & \.0.3331 \\
H(3)        & -0.4673     & -0.8095     & -0.3164     & ---      & ---           & \.0.3331 \\ \hline\hline
\end{tabular}
\end{center}
\end{table}

\newpage
\begin{table}[h]
\noindent
\caption{Vapor-liquid equilibria of ammonia: simulation results using the new model (sim) compared to data from a reference quality equation of state \cite{Tillner-Roth1993} (eos) for vapor pressure, saturated densities and enthalpy of vaporization. The number in parentheses indicates the statistical uncertainty in the last digit.}
\label{tab_vle}

\medskip
\begin{center}
\begin{tabular}{c|cccccccc} \hline\hline
$T$ & $p_{\mathrm{sim}}$ & $p_{\mathrm{eos}}$ & $\rho'_{\mathrm{sim}}$ & $\rho'_{\mathrm{eos}}$ & $\rho''_{\mathrm{sim}}$ & $\rho''_{\mathrm{eos}}$ & $\Delta h^{\mathrm{v}}_{\mathrm{sim}}$ & $\Delta h^{\mathrm{v}}_{\mathrm{eos}}$ \\
K & MPa & MPa & mol/l & mol/l & mol/l & mol/l & kJ/mol & kJ/mol \\ \hline
240 & 0.12(1) & 0.102 & 40.26(1)  & 40.032 & 0.066(5)   & 0.0527  & 24.11(1)   & 23.31 \\
280 & 0.60(2) & 0.551 & 36.98(2)  & 36.939 & 0.280(8)   & 0.257\0 & 21.56(1)   & 21.07 \\
315 & 1.65(4) & 1.637 & 33.76(3)  & 33.848 & 0.74\0(1)  & 0.744\0 & 18.96(2)   & 18.57 \\
345 & 3.37(4) & 3.457 & 30.45(4)  & 30.688 & 1.55\0(1)  & 1.624\0 & 16.19(3)   & 15.79 \\
363 & 5.22(5) & 5.101 & 28.17(6)  & 28.368 & 2.56\0(2) & 2.544\0 & 13.93(5)   & 13.65 \\
375 & 6.37(6) & 6.485 & 26.18(7)  & 26.502 & 3.17\0(3) & 3.459\0 & 12.48(6)   & 11.89 \\
385 & 7.88(5) & 7.845 & 24.05(9)  & 24.608 & 4.27\0(5) & 4.554\0 & 10.49(9)   & 10.08 \\
395 & 9.54(7) & 9.422 & 20.9\0(1) & 22.090 & 5.66\0(9) & 6.272\0 & \08.1\0(1) & \07.66 \\ \hline\hline
\end{tabular}
\end{center}
\end{table}

% figures
\newpage
\listoffigures

\newpage

\newpage

\newpage

\newpage

\newpage

\newpage

\newpage

\newpage


\begin{thebibliography}{00}

\bibitem{Ungerer2006}
P. Ungerer, V. Lachet, and B. Tavitian.
\newblock Applications of molecular simulation in oil and gas production and processing.
\newblock \emph{Oil Gas Sci. Technol.}, 61 (2006) 387--403.

\bibitem{Eckl2008}
B. Eckl, J. Vrabec, and H. Hasse.
\newblock On the application of force fields for predicting a wide variety of properties:
Ethylene oxide as an example.
\newblock \emph{Fluid Phase Equilib.}, submitted (2007).

\bibitem{Eckl2007}
B. Eckl, Y.-L.Huang, J. Vrabec, and H. Hasse.
\newblock Vapor pressure of R227ea + Ethanol at 343.17 K by molecular simulation.
\newblock \emph{Fluid Phase Equilib.}, 260 (2007) 177--182.

\bibitem{Jorgensen1980}
W.L. Jorgensen and M. Ibrahim.
\newblock Structure and properties of liquid ammonia.
\newblock \emph{J. Amer. Chem. Soc.} 102 (1980) 3309--3315.

\bibitem{Hinchliffe1981}
A. Hinchliffe, D.G. Bounds, M.L. Klein, I.R. McDonald, and R. Righini.
\newblock Intermolecular potentials for ammonia based on SCF-MO calculations.
\newblock \emph{J. Chem. Phys.} 74 (1981) 1211--1216.

\bibitem{Impey1984}
R.W. Impey and M.L. Klein.
\newblock A simple intermolecular potential for liquid ammonia.
\newblock \emph{Chem. Phys. Lett.} 104 (1984) 579--582.

\bibitem{Narten1977}
A.H. Narten.
\newblock Liquid-ammonia - molecular correlation-functions from x-ray-diffraction.
\newblock \emph{J. Chem. Phys.} 66 (1977) 3117--3120.

\bibitem{Kristof1999}
T. Krist\'of, J. Vorholz, J. Liszi, B. Rumpf, and G. Maurer.
\newblock A simple effective pair potential for the molecular simulation of the thermodynamic properties of ammonia.
\newblock \emph{Mol. Phys.} 97 (1999) 1129--1137.

\bibitem{Pana1987}
A.Z. Panagiotopoulos.
\newblock Direct determination of phase coexistence properties of fluids by Monte-Carlo simulation in a new ensemble.
\newblock \emph{Mol. Phys.}, 61 (1987) 813--826.

\bibitem{Pana1988}
A.Z. Panagiotopoulos, N. Quirke, M. Stapleton, and D.J. Tildesley.
\newblock Phase-equilibria by simulation in the Gibbs ensemble - alternative derivation, generalization and application to mixture and membrane equilibria.
\newblock \emph{Mol. Phys.} 63 (1988) 527--545.

\bibitem{Kristof1996}
T. Krist\'of, J. Liszi.
\newblock Alternative implementations of the Gibbs ensemble Monte Carlo calculation.
\newblock \emph{Chem. Phys. Lett.}, 261 (1996) 620--624.

\bibitem{Kristof1997}
T. Krist\'of, J. Liszi.
\newblock Application of a new Gibbs ensemble Monte Carlo method to site-site interaction model fluids.
\newblock \emph{Mol. Phys.}, 90 (1997) 1031--1034.

\bibitem{Schnabel2007}
T. Schnabel, M. Cortada, J. Vrabec, S. Lago, and H. Hasse.
\newblock Molecular model for formic acid adjusted to vapor-liquid equilibria.
\newblock \emph{Chem. Phys. Lett.}, 435 (2007) 268--272.

\bibitem{Schnabel2005}
T. Schnabel, J. Vrabec, and H. Hasse.
\newblock Henry's law constants of methane, nitrogen, oxygen and carbon dioxide in ethanol from 273 to 498 K: prediction from molecular simulation.
\newblock \emph{Fluid Phase Equilib.}, 233 (2005) 134--143.

\bibitem{Schnabel2007a}
T. Schnabel, A. Srivastava, J. Vrabec, and H. Hasse.
\newblock Hydrogen Bonding of Methanol in Supercritical CO2: Comparison between 1H-NMR Spectroscopic Data and Molecular Simulation Results,
\newblock \emph{J. Phys. Chem. B}, 111 (2007) 9871--9878.

\bibitem{Schmidt1993}
M.W. Schmidt, K.K. Baldridge, J.A. Boatz, S.T. Elbert, M.S. Gordon, J.H. Jensen, S. Koseki, N. Matsunaga, K.A. Nguyen, Kiet.
\newblock General atomic and molecular electronic structure system.
\newblock \emph{J. Comput. Chem.}, 14 (1993) 1347--1363.

\bibitem{Vrabec2001}
J. Vrabec, J. Stoll, and H. Hasse.
\newblock A set of molecular models for symmetric quadrupolar fluids.
\newblock \emph{J. Phys. Chem. B}, 105 (2001) 12126--12133.

\bibitem{Stoll2003}
J. Stoll, J. Vrabec, and H. Hasse.
\newblock A set of molecular models for carbon monoxide and halogenated hydrocarbons.
\newblock \emph{J. Chem. Phys.}, 119 (2003) 11396--11407.

\bibitem{Baldridge1997}
K. Baldridge and A. Klamt.
\newblock First principles implementation of solvent effects without outlying charge error.
\newblock \emph{J. Chem. Phys.}, 106 (1997) 6622--6633.

\bibitem{Stoll2005}
J. Stoll.
\newblock \emph{Molecular Models for the Prediction of Thermophysical Properties of Pure Fluids and Mixtures}.
\newblock Fortschritt-Berichte VDI, Reihe 3, 836, VDI-Verlag, D\"usseldorf, 2005.

\bibitem{Ungerer1999}
P. Ungerer, A. Boutin, and A.H. Fuchs.
\newblock Direct calculation of bubble points by Monte Carlo simulation.
\newblock \emph{Mol. Phys.}, 97 (1999) 523--539.

\bibitem{Bourasseau2003}
E. Bourasseau, M. Haboudou, A. Boutin, A.H. Fuchs, and P. Ungerer.
\newblock New optimization method for intermolecular potentials: Optimization of a new anisotropic united atoms potential for olefins: Prediction of equilibrium properties.
\newblock \emph{J. Chem. Phys.}, 118 (2003) 3020--3034.

\bibitem{Tillner-Roth1993}
R. Tillner-Roth, F. Harms-Watzenberg, and H. D. Baehr.
\newblock Eine neue Fundamentalgleichung fuer Ammoniak.
\newblock \emph{DKV-Tagungsbericht}, 20 (1993) 167--181.

\bibitem{Haar1978}
L. Haar.
\newblock Thermodynamic properties of ammonia.
\newblock \emph{J. Phys. Chem. Ref. Data}, 7 (1978) 635--792.

\bibitem{Ahrendts1979}
J. Ahrendts and H.D. Baehr.
\newblock \emph{Die thermodynamischen Eigenschaften von Ammoniak}.
\newblock VDI-Forschungsheft Nr. 596, VDI-Verlag, D\"usseldorf, 1979.

\bibitem{REFPROP}
E.W. Lemmon, M.L. Huber, and M.O. McLinden.
\newblock \emph{REFPROP, NIST Standard Reference Database 23, Version 8.0}.
\newblock Physical and Chemical Properties Division, National Institute of Standards and Technology, Boulder, 200x.

\bibitem{Mathews1972}
J.F. Mathews.
\newblock The critical constants of inorganic substances.
\newblock \emph{Chem. Rev.}, 72 (1972) 71--100.

\bibitem{Lotfi1992}
A. Lotfi, J. Vrabec, and J. Fischer,
\newblock Vapour liquid equilibria of the Lennard-Jones fluid from the $NpT$ plus test particle method.
\newblock \emph{Mol. Phys.} 76 (1992) 1319--1333.

\bibitem{Mayer1937}
J.E. Mayer.
\newblock The statistical mechanics of condensing systems. I.
\newblock \emph{J. Chem. Phys.}, 5 (1937) 67--73.

\bibitem{Mayer1939}
J.E. Mayer.
\newblock Statistical mechanics of condensing systems V Two-component systems.
\newblock \emph{J. Phys. Chem.}, 43 (1939) 71--95.

\bibitem{Mayer1940}
J.E. Mayer and M.G. Mayer.
\newblock \emph{Statistical Mechanics}, John Wiley and Sons, New York, 1940.

\bibitem{Gray1984}
C. G. Gray, K. E. Gubbins.
\newblock \emph{Theory of molecular fluids, 1. Fundamentals}, Clarendon Press, Oxford, 1984.

\bibitem{Mountain2005}
R.D. Mountain.
\newblock A polarizable model for ethylene oxide.
\newblock \emph{J. Phys. Chem. B}, 109 (2005) 13352--13355.

\bibitem{Allen1987}
M. P. Allen, D. J. Tildesley.
\newblock \emph{Computer simulations of liquids.} 
Clarendon Press, Oxford, 1987.

\bibitem{Ricci1995}
M. Ricci, M. Nardone, F. Ricci, C. Andreani, and A. Soper.
\newblock Microscopic structure of low temperature liquid ammonia: A neutron diffraction experiment.
\newblock \emph{J. Chem. Phys.} 102 (1995) 7650--7655.

\bibitem{Vrabec2002}
J. Vrabec and H. Hasse.
\newblock Grand Equilibrium: vapour-liquid equilibria by a new molecular simulation method.
\newblock \emph{Mol. Phys.}, 100 (2002) 3375--3383.

\bibitem{Anderson1980}
H. C. Andersen.
\newblock Molecular dynamics simulations at constant pressure and/or temperature.
\newblock \emph{J. Chem. Phys.}, 72 (1980) 2384--2393.

\bibitem{Widom1963}
B. Widom.
\newblock Some topics in the theory of fluids.
\newblock \emph{J. Chem. Phys.}, 39 (1963) 2808--2812.

\bibitem{Nezbeda1991}
I. Nezbeda, J. Kolafa.
\newblock A New Version of the Insertion Particle Method for Determining the Chemical Potential by Monte Carlo Simulation.
\newblock \emph{Mol. Sim.}, 5 (1991) 391--403.

\bibitem{Vrabec2002a}
J. Vrabec, M. Kettler, and H. Hasse.
\newblock Chemical potential of quadrupolar two-centre Lennard-Jones fluids by gradual insertion.
\newblock \emph{Chem. Phys. Lett.}, 356 (2002) 431--436.

\bibitem{Flyvbjerg1989} 
H. Flyvbjerg, H. G. Petersen.
\newblock Error estimates on averages of correlated data.
\newblock \emph{J. Chem. Phys.}, 91 (1989) 461--466.

\end{thebibliography}
\end{document}